\documentclass[12pt,english]{paper}
\usepackage{amssymb}
\usepackage[T1]{fontenc}
\usepackage[latin1]{inputenc}
\usepackage{graphicx}
\usepackage{cite}
\makeatletter

\newcommand{\bibstyle@aas}{\bibpunct{(}{)}{;}{a}{}{,}}

\makeatletter

\makeatletter

\usepackage{geometry}

\makeatletter

\usepackage{epsfig}

\makeatother

\makeatother

\makeatother

\usepackage{babel}
\makeatother
\begin{document}

\title{Direct detection of Ultra-High Energy WIMPs with a satellite detector like JEM-EUSO}

\author{Ye Xu$^{1,2}$}

\maketitle

\begin{flushleft}
$^1$School of Information Science and Engineering,  Fujian University of Technology, Fuzhou 350118, China
\par
$^2$Research center for Microelectronics Technology, Fujian University of Technology, Fuzhou 350118, China
\par
e-mail address: xuy@fjut.edu.cn
\end{flushleft}

\begin{abstract}
The possibility of directly detecting Ultra-high energy (UHE from now on) WIMPs are considered by the WIMPs interaction with the nuclei in the air. Since neutrinos dominate the events from the spherical crown near JEM-EUSO, all the events from this region are ignored in my work. Then the numbers of UHE WIMPs and neutrino detected by JEM-EUSO are evaluated at different energies (1 PeV < E < 100 EeV) in ten years, respectively. If the energy thresholds are taken to be 20 EeV, neutrino events can be almost rejected in the detection of UHE WIMPs. According to my evaluation, O(10-100) UHE WIMP events can be detected by JEM-EUSO at energies above 20 EeV in ten years.
\end{abstract}


\section{Introduction}
It is indicated by the Planck data with measurements of the cosmic microwave background
 that $26.6\%$ of the overall energy density of
the Universe is non-baryonic dark matter\cite{Planck2015}. Weakly Interacting Massive Particles (WIMPs from now on),
predicted by extensions of the Standard Model of particle physics,
are a class of candidates for dark matter\cite{GDJ}. They are
distributed in a halo surrounding a galaxy. A WIMP halo of a galaxy
with a local density of 0.3 GeV/cm$^3$ is assumed and its relative
speed to the Sun is 230 km/s\cite{JP}. At present, one mainly searches for thermal
WIMPs via direct and indirect detections\cite{CDMSII,CDEX,XENON1T,LUX,PANDAX,AMS-02,DAMPE,fermi}. Because of the very small cross sections of the interactions
between these WIMPs and nucleus (maybe O(10$^{-47}$ cm$^2$))\cite{XENON1T,PANDAX}, so far one has not found dark matter yet.
\par
It is a reasonable assumption that there exist various dark matter particles in the Universe. Then it is possible that
this sector may comprise of non-thermal (and non-relativistic) components. And these particles may also contain a small
component which is relativistic and highly energetic. Although the fraction of these relativistic dark matter particles
is small in the Universe, their large interaction cross sections (including between themselves and between them and the
Standard Model (SM from now on) particles) make it possible to find them. Due to the reasons mentioned above, one has
to shift more attention to direct and indirect detection of UHE WIMPs. In fact, A. Bhattecharya, R. Gaudhi and A. Gupta have discussed the possibility that the PeV events are UHE dark matter particles at IceCube in their work\cite{BGA}. And it has been discussed the possibility that UHE WIMPs are indirectly probed by IceCube via detecting UHE neutrino signatures from the earth core in my other work\cite{xu1}.
\par
The relativistic WIMPs are mainly generated by two mechanisms in the Universe. One is through the collision of UHE cosmic ray particles and
thermal WIMPs. This collision will result in some UHE WIMPs flux. The other is that UHE WIMPs can also originate from
the early Universe. There are a non-thermal dark sector generated by the early Universe with its bulk comprised of a very massive
relic $\phi$ in the Universe. This superheavy dark matter\cite{KC87, CKR98, CKR99, CGIT, FKMY} decays to another much lighter WIMPs $\chi$ and its lifetime is greater
than the age of the Universe. This lead to a small but significant flux of UHE WIMPs\cite{LT, BGA, EIP, BLS, BKMTZ}.
\par
In the present work, it is only focused on direct detection of the UHE WIMPs $\chi$ induced by the decay of superheavy dark matter $\phi$ ($\phi\to\chi\bar{\chi}$). These UHE WIMPs $\chi$, which pass through the Earth and air and interact with nuclei, can be detected by a satellite detector like JEM-EUSO\cite{JEM-EUSO}, via fluorescent and Cherenkov photons due to the development of extensive air showers. In this detection, the main contamination is from the diffuse neutrinos in the Galaxy.
\par
In what follows, the UHE WIMP and background event rates from diffuse neutrinos will be estimated at JEM-EUSO. And It is discussed the possibility of direct detection of UHE WIMPs induced by the decay of superheavy dark matter.
\section{UHE WIMPs flux in the Galaxy}
It is considered a scenario where the dark matter sector is composed of at least two particle species in the Universe\cite{BGA}. One is a co-moving non-relativistic scalar species $\phi$, with mass $m_{\phi} \geq$ 1 PeV, the other is much lighter particle species $\chi$ ($m_{\chi} \ll m_{\phi}$), due to the decay of $\phi$, with a very large lifetime. And $\chi$ comprises the bulk of present-day dark matter. The lifetime for the decay of heavy dark matter to standard model (SM from now on) particles is strongly constrained ($\tau \geq$ O($10^{26}-10^{29}$)s) by diffuse gamma and neutrino observations\cite{EIP,MB,RKP,KKK}. In the present work, it is considered an assumption that superheavy dark matter could not decay to SM particles. And $\tau_{\phi}$ is taken to be $10^{25}$s. Since the WIMP flux only depends on the two-body decay of superheavy dark matter and the dark matter distribution in the galactic halo, it is the same as the neutrino flux due to the decay of superheavy dark matter in Ref.\cite{EIP,BLS}:
\begin{center}
\begin{equation}
\psi_{\chi}=\displaystyle\frac{1}{4\pi m_{\phi}\tau_{\phi}}\int \displaystyle\frac{dN_{\chi}}{dE_{\chi}}\rho_{halo}dsdE
\end{equation}
\end{center}
where $\rho_{halo}$ is the density profile of dark matter particles in the Galaxy and s is a line-of-sight. $\displaystyle\frac{dN_{\chi}}{dE_{\chi}}=\delta(E_{\chi}-\displaystyle\frac{m_{\phi}}{2})$, and E$_{\chi}$ and N$_{\chi}$ are the energy and number of UHE WIMP, respectively.
Then the UHE WIMPs flux from the Galaxy is obtained via the following equation\cite{BLS}:
\begin{center}
\begin{equation}
\psi_{\chi}=\int_{E_{min}}^{E_{max}}F\frac{dN_\chi}{dE_\chi}dE
\end{equation}
\end{center}
with
\par
\begin{center}
\begin{equation}
F=1.7\times10^{-12}cm^{-2}s^{-1}\times\frac{10^{28}s}{\tau_{\phi}}\times\frac{1PeV}{m_{\phi}}.
\end{equation}
\end{center}
\section{UHE WIMP interaction with nuclei}
In the present paper, a Z' portal dark matter model\cite{APQ,Hooper} is taken for WIMPs to interact with nuclei within the JEM-EUSO detecting zone (see Fig.1). In this model, a new Z' gauge boson is considered as a simple and well-motivated extension of SM (see Fig.1(a) in Ref.\cite{BGA}). And the parameters in the model are taken to be the same as the ones in Ref.\cite{BGA}, that is, the interaction vertexes ($\chi\chi$Z' and qqZ') are assumed to be vector-like, the coupling constant G ($G=g_{\chi\chi Z}g_{qqZ}$) is chosen to be 0.05 and the Z' and $\chi$ masses are taken to be 5 TeV, 10 GeV, respectively. Theoretical models that encompass WIMP spectrum have been discussed in the literature in terms of Z or Z' portal sectors with Z' vector boson typically acquiring mass through the breaking of an additional U(1) gauge group at the high energies (see Ref.\cite{APQ,Hooper}). The UHE WIMP interaction cross section with nucleus is obtained by the following function(see Fig.1(b) in Ref.\cite{BGA}):
\begin{center}
\begin{equation}
\sigma_{\chi N}=6.13\times10^{-43} cm^2 \left(\frac{E_{\chi}}{1GeV}\right)^{0.518}
\end{equation}
\end{center}
where E$_{\chi}$ is the UHE WIMP energy.
\par
For neutrino energies above 1 PeV, the interaction cross sections with nucleus are given by simple power-law forms\cite{BHM}:
\begin{center}
\begin{equation}
\sigma_{\nu N}(CC)=4.74\times10^{-35} cm^2 \left(\frac{E_{\nu}}{1 GeV}\right)^{0.251}
\end{equation}
\end{center}
\begin{center}
\begin{equation}
\sigma_{\nu N}(NC)=1.80\times10^{-35} cm^2 \left(\frac{E_{\nu}}{1 GeV}\right)^{0.256}
\end{equation}
\end{center}
where $E_{\nu}$ is the neutrino energy.Then the above equations show that $\sigma_{\chi N}$ is smaller by 10-11 orders of magnitude, compared to $\sigma_{\nu,\bar{\nu} N}$, at energies between 20EeV and 100EeV.
\par
The WIMP and neutrino interaction length can be obtained by
\par
\begin{center}
\begin{equation}
L_{\nu,\chi}=\frac{1}{N_A\rho\sigma_{\nu,\chi N}}
\end{equation}
\end{center}
\par
where $N_A$ is the Avogadro constant, and $\rho$ is the density of matter, which WIMPs and neutrinos interact with.
\section{Evaluation of the numbers of UHE WIMPs and neutrinos detected by JEM-EUSO}
UHE WIMPs reach the Earth and pass through the Earth and air, meanwhile these particles interact with matter of the Earth and air. Hadrons are produced by UHE WIMP interaction with atmospheric nuclei. The secondary particles generated by these UHE hadrons will develop into a cascade. And the most dominant particles in a cascade are electrons moving through atmosphere. Ultraviolet fluorescence photons are emitted by electron interaction with nitrogen. The emitted photons are isotropic and their intensity is proportional to the energy deposited in the atmosphere. A small part of them will be detected by the JEM-EUSO detector (see Fig. 1). Since these signatures are similar to deep inelastic scattering of UHE neutrinos via neutral current, JEM-EUSO is unable to discriminate between their signatures. Only the geometrical analysis is used to discriminate between UHE WIMPs and neutrinos in the present paper.
\par
JEM-EUSO is a space science observatory to explore the extreme-energy cosmic rays and upward-going neutrinos in the Universe\cite{eusomission}. It will be installed into the International Space Station (ISS from now on) in 2019. The ISS maintains an orbit with an altitude of $\sim$400 km and circles the Earth in roughly 90 minutes. The JEM-EUSO telescope has a wide field of view (FOV: $\pm30^{\circ}$) and observes extreme energy particles in the two modes (nadir and tilted modes) via fluorescent and Cherenkov photons due to the development of extensive air showers. JEM-EUSO is tilted by an angle of 30 degrees in the tilted mode. JEM-EUSO has a observational area of about $2\times10^5$ km$^2$ and $7\times10^5$ km$^2$ in nadir and tilted modes, respectively. The duty cycle for JEM-EUSO, R, is taken to be 10\%.
\par
The number of UHE WIMPs, N$_{det}$, detected by JEM-EUSO can be obtained by the following function:
\begin{center}
\begin{equation}
N_{det} = R\times T\times (A\Omega)_{eff} \times \Phi_{\chi}
\end{equation}
\end{center}
where T is the lifetime of the JEM-EUSO experiment, $\Phi_\chi=\displaystyle\frac{d\psi_\chi}{dE_{\chi}}$ and $(A\Omega)_{eff}$ = the observational area $\times$ the effective solid angle $\times$ P(E,$D_e$,D). Here $P(E,D_e,D)=exp\left(-\displaystyle\frac{D_e}{L_{earth}}\right)\left[1-exp\left(-\displaystyle\frac{D}{L_{air}}\right)\right]$ is the probability that UHE WIMPs interacts with air after traveling a distance between $D_e$ and $D_e+D$, where D is the effective length in the JEM-EUSO detecting zone in the air, $D_e$ are the distances through the Earth, and $L_{earth,air}$ are the UHE WIMP interaction lengths with the Earth and air, respectively.
\par
In what follows, $(A\Omega)_{eff}$ is roughly considered and then the numbers of UHE WIMPs detected by JEM-EUSO are evaluated. Here it is made the assumption that the observational area of JEM-EUSO is regarded as a point in the calculation of the effective solid angle $\Omega$. Under this approximation,
\begin{center}
\begin{equation}
(A\Omega)_{eff} \approx P(E,D_e,D)A\int_0^{\theta_{max}} \frac{2\pi {R_e}^2sin\theta}{D_e^2} d\theta .
\end{equation}
\end{center}
where A is the observational area, $R_e$ is the radius of the Earth, $\theta$ is the polar angle for the Earth (see Fig. 1), $\theta_{max}$ is the maximum of $\theta$, $D_e=\displaystyle\frac{R_e(1+cos\theta)}{cos\displaystyle\frac{\theta}{2}}$. $D=\displaystyle\frac{H}{ctg30^{\circ}sin\displaystyle\frac{\theta}{2}+cos\displaystyle\frac{\theta}{2}}$ and $D=H(cos\displaystyle\frac{\theta}{2}+sin\displaystyle\frac{\theta}{2}ctan(\displaystyle\frac{2\pi}{3}-\displaystyle\frac{\theta}{2}))$ in the nadir and tilted modes, respectively. Here H is the altitude of ISS.
\par
The background due to diffuse neutrinos is roughly estimated with a diffuse neutrino flux of $\Phi_{\nu}=0.9^{+0.30}_{-0.27}\times(E_{\nu}/100TeV)\times10^{-18}GeV^{-1} cm^{-2}s^{-1}sr^{-1}$\cite{icecube}, where $\Phi_{\nu}$ represents the per-flavor flux, by the above method.

\section{Results}
The numbers of UHE WIMPs and neutrino detected by JEM-EUSO are evaluated at different energy at different $\theta$, respectively. We can obtain neutrino contamination percentages in the upward-going events detected by JEM-EUSO by
\begin{center}
\begin{equation}
Neutrino\% = \frac{N_{\nu}}{N_{WIMP}+N_{\nu}}
\end{equation}
\end{center}
where $N_{WIMP}$ and $N_{\nu}$ are the numbers of UHE WIMPs and neutrinos detected by JEM-EUSO at different energy at different $\theta$, respectively. Fig. 2 shows Neutrino\% at different energy at different $\theta$ when $\tau_\phi=10^{25}$s. From this figure, we can know neutrinos dominate the detected upward-going events in the spherical crown near the observational area of JEM-EUSO at energies below $\sim$100 PeV. So the events from this zone should be ignored for rejecting neutrino background. Besides, this figure shows the neutrino contamination is less than $10^{-10}$ in the signal region of 0<$\theta$<150$^\circ$ and 20 EeV<E<100EeV. So we almost confirm the events are UHE WIMPs from this region.
\subsection{Nadir mode}
The numbers of UHE WIMPs and neutrino detected by JEM-EUSO are evaluated at different energy (1 PeV < E < 100 EeV) at the different $\theta_{max}$ ($150^{\circ},120^{\circ},$ and $90^{\circ}$) in ten years in the nadir mode when $\tau_\phi=10^{25}$s, respectively (see Fig.3-5). Fig. 3 shows that the numbers of UHE WIMPs and neutrinos at $\theta_{max}=150^{\circ}$. If the energy threshold for JEM-EUSO is taken into account (about 20 EeV\cite{threshold}), the numbers of the detected UHE WIMPs can reach about 50 and 23 at at the energies with 20 EeV and 100 EeV in ten years, respectively. Fig. 4 shows that the numbers of UHE WIMPs and neutrinos at $\theta_{max}=120^{\circ}$. The numbers of the detected UHE WIMPs can reach about 27 and 12 at at the energies with 20 EeV and 100 EeV in ten years, respectively. Fig. 5 shows that the numbers of UHE WIMPs and neutrinos at $\theta_{max}=90^{\circ}$, respectively. The numbers of the detected UHE WIMPs can reach about 14 and 7 at the energies with 20 EeV and 100 EeV in ten years, respectively.
\subsection{Tilted mode}
The numbers of UHE WIMPs and neutrino detected by JEM-EUSO are evaluated at different energy (1 PeV < E < 100 EeV) at the different $\theta_{max}$ ($150^{\circ},120^{\circ},$ and $90^{\circ}$) in ten years in the tilted mode when $\tau_\phi=10^{25}$s, respectively (see Fig.6-8). Fig. 6 shows that the numbers of UHE WIMPs and neutrinos at $\theta_{max}=150^{\circ}$. If the energy threshold for JEM-EUSO is taken into account (about 20 EeV\cite{threshold}), the numbers of the detected UHE WIMPs can reach about 340 and 157 at at the energies with 20 EeV and 100 EeV in ten years, respectively. Fig. 7 shows that the numbers of UHE WIMPs and neutrinos at $\theta_{max}=120^{\circ}$. The numbers of the detected UHE WIMPs can reach about 157 and 72 at at the energies with 20 EeV and 100 EeV in ten years, respectively. Fig. 8 shows that the numbers of UHE WIMPs and neutrinos at $\theta_{max}=90^{\circ}$, respectively. The numbers of the detected UHE WIMPs can reach about 75 and 35 at the energies with 20 EeV and 100 EeV in ten years, respectively.
\section{Discussion and conclusion}
According to the results described above, it is possible that UHE WIMPs are directly detected with a satellite detector like JEM-EUSO. It is made an approximation that the observational area is regards as a point in the calculation of the solid angle. This produces some deviations for the event rates of UHE WIMPs and neutrinos, but they can not have an effect on the conclusion that O(10-100) UHE WIMP events can be detected by JEM-EUSO in ten years, especially, $\theta_{max}$ is a less value, such as $\theta_{max} \le 120^{\circ}$.
\par
Constrained by the lifetime of ISS, JEM-EUSO has the operation time of five years. So the WIMP event rates have only a half of the above evaluated ones. Since $\Phi_{\chi}$ is proportional to $\displaystyle\frac{1}{\tau_{\phi}}$, the above results are actually depended on the lifetime of superheavy dark matter. For example, the WIMP event rate for JEM-EUSO is $\sim$800 and $\sim$400 events/five years at 20EeV and 100EeV in the case of $\theta_{max}=120^{\circ}$ in the tilted mode when $\tau_{\phi}=10^{24}$s, respectively. The WIMP event rate for JEM-EUSO is $\sim$80 and $\sim$40 events/five years at 20EeV and 100EeV in the case of $\theta_{max}=120^{\circ}$ in the tilted mode when $\tau_{\phi}=10^{24}$s, respectively. And the WIMP event rate for JEM-EUSO is $\sim$8 and $\sim$4 events/five years at 20EeV and 100EeV in the case of $\theta_{max}=120^{\circ}$ in the tilted mode when $\tau_{\phi}=10^{26}$s, respectively.
\par
Thus it can be seen, it is possible that UHE WIMPs are detected at JEM-EUSO under the assumption that the superheavy dark matter $\phi$ is unable to decay to SM particles. If the signatures from the signal region (see Fig. 2) are measured by JEM-EUSO, there must be new physical particles, which could be UHE WIMPs, in the Universe. It is also possible that UHE WIMPs are directly probed by the detectors based on ground, such as the Pierre Auger observatory, when the lifetime of superheavy dark matter is less than $10^{24}$s. Besides, UHE SM particles due to the annihilation between UHE and thermal WIMPs could be measured. For example, UHE neutrinos from the solar center due to the annihilation between UHE and thermal WIMPs could be measured by a neutrino telescope like IceCube.
\section{Acknowledgements}
This work was supported by the National Natural Science Foundation
of China (NSFC) under the contract No. 11235006, the Science Fund of
Fujian University of Technology under the contract No. GY-Z14061 and the Natural Science Foundation of
Fujian Province in China under the contract No. 2015J01577.
\par

\newpage

\begin{figure}
 \centering
 \includegraphics[width=0.9\textwidth]{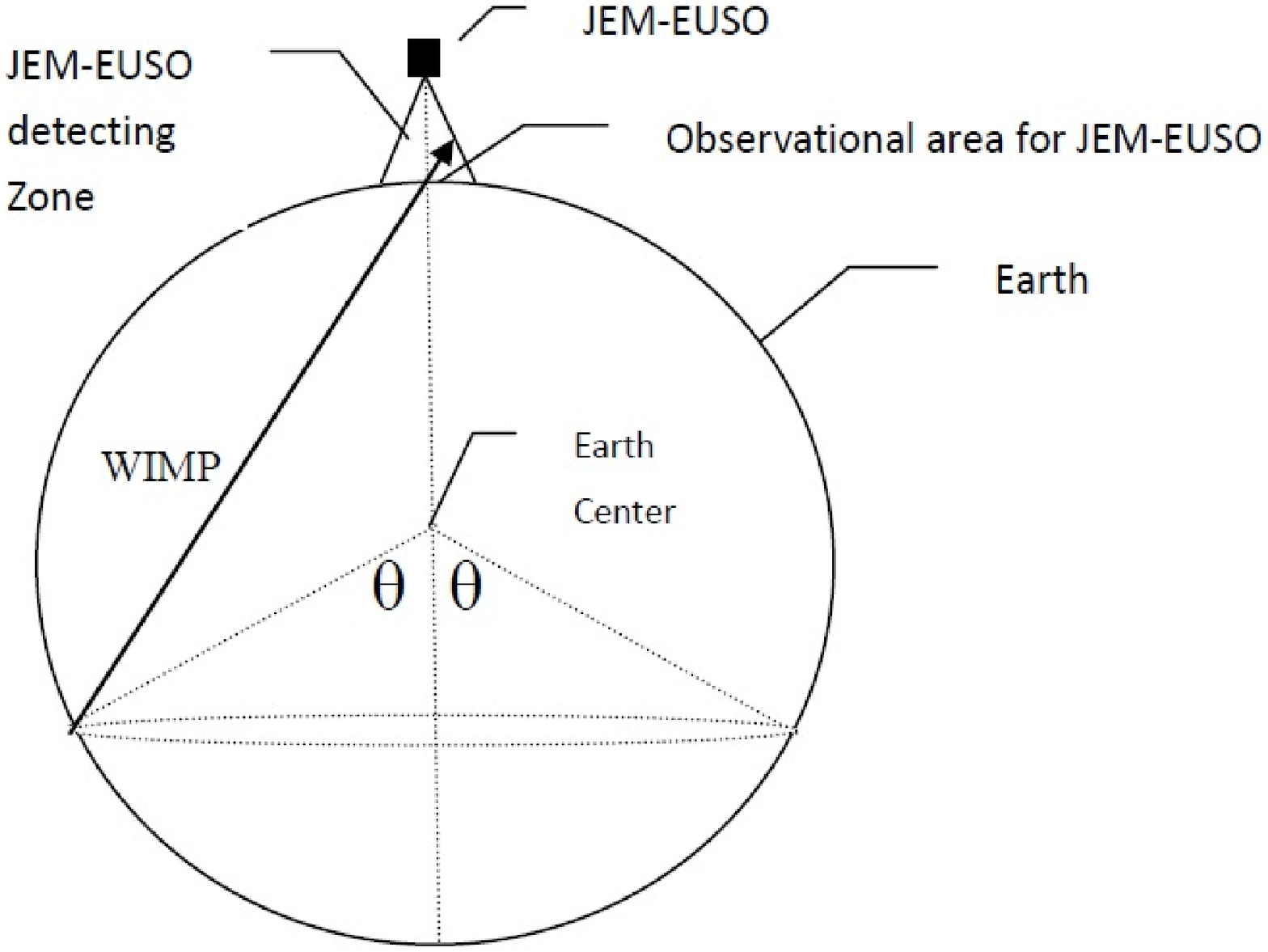}
 \caption{UHE WIMPs pass through the Earth and air and can be detected by a satellite detector like JEM-EUSO, via fluorescent and Cherenkov photons due to the development of extensive air showers. $\theta$ is the polar angle for the Earth.}
 \label{fig:figure}
\end{figure}

\begin{figure}
 \centering
 \includegraphics[width=0.9\textwidth]{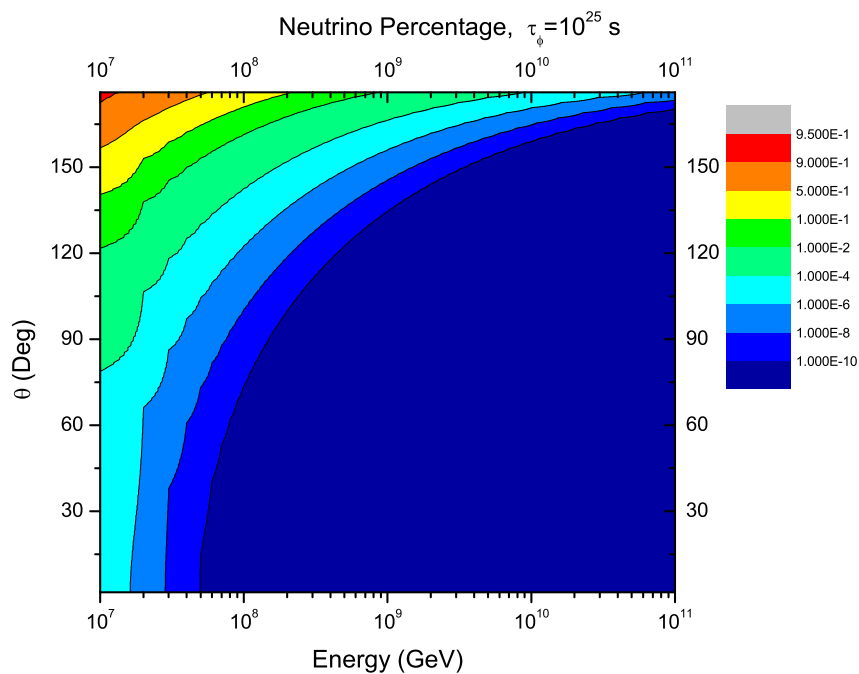}
 \caption{Neutrino contamination percentages are evaluated in the upward-going events detected by JEM-EUSO when $\tau_{\phi}=10^{25}$s}
 \label{fig:nu_per_25}
\end{figure}

\begin{figure}
 \centering
 \includegraphics[width=0.9\textwidth]{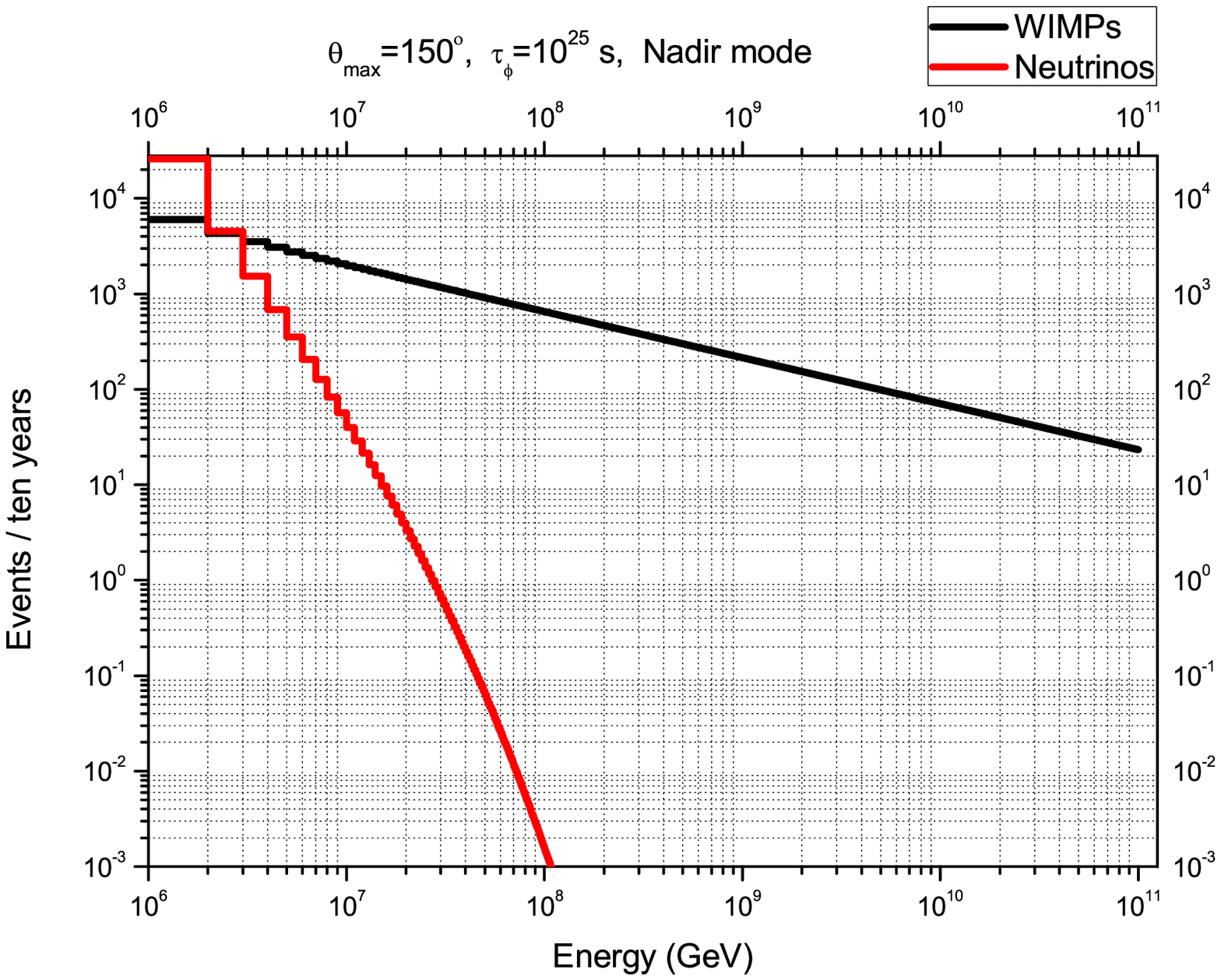}
 \caption{The WIMP and neutrino event rates are evaluated at $\theta_{max} = 150^{\circ}$ in the nadir mode at JEM-EUSO when $\tau_{\phi}=10^{25}$s}
 \label{fig:150_25}
\end{figure}

\begin{figure}
 \centering
 \includegraphics[width=0.9\textwidth]{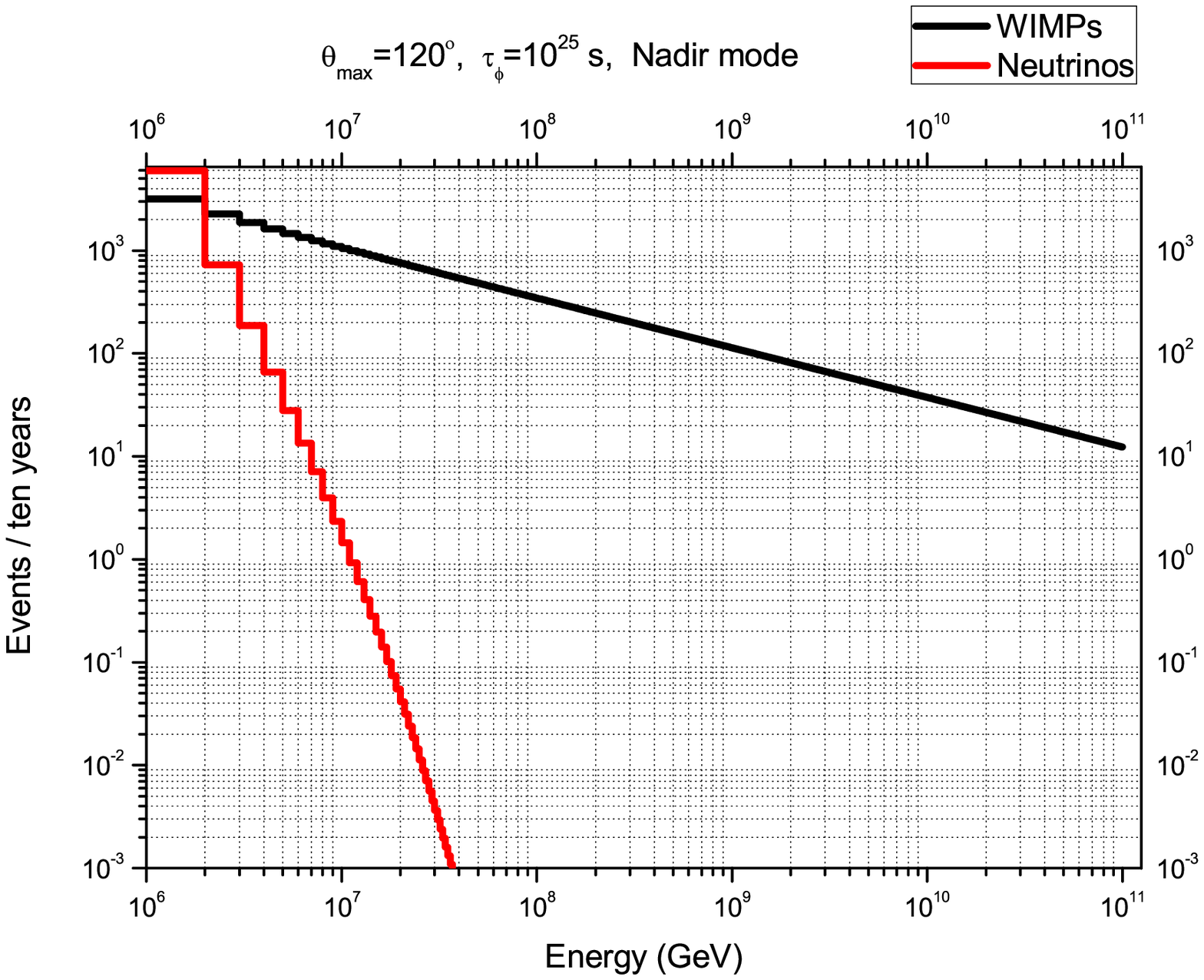}
 \caption{The WIMP and neutrino event rates are evaluated at $\theta_{max} = 120^{\circ}$ in the nadir mode at JEM-EUSO when $\tau_{\phi}=10^{25}$s}
 \label{fig:120_25}
\end{figure}

\begin{figure}
 \centering
 \includegraphics[width=0.9\textwidth]{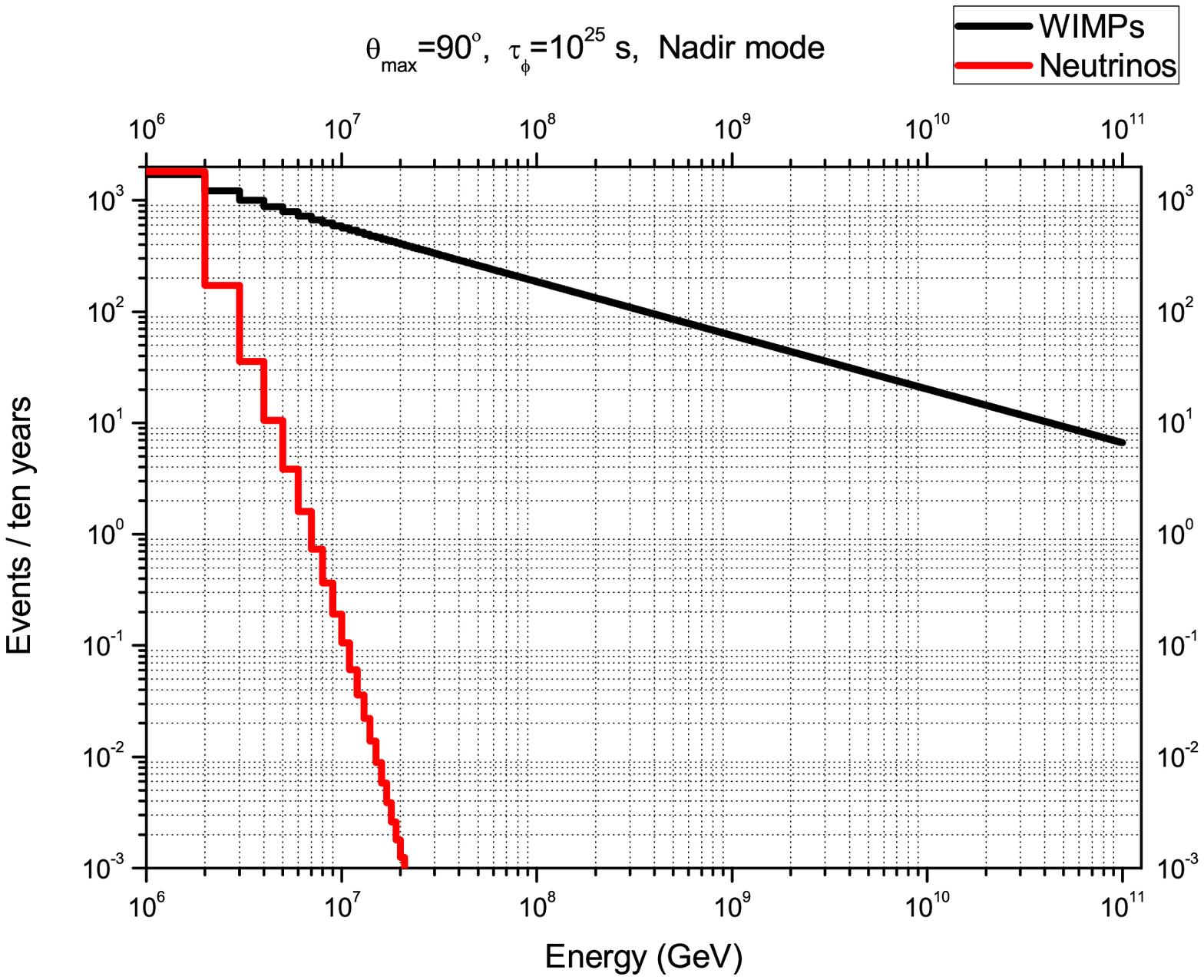}
 \caption{The WIMP and neutrino event rates are evaluated at $\theta_{max} = 90^{\circ}$ in the nadir mode at JEM-EUSO when $\tau_{\phi}=10^{25}$s}
 \label{fig:90_25}
\end{figure}

\begin{figure}
 \centering
 \includegraphics[width=0.9\textwidth]{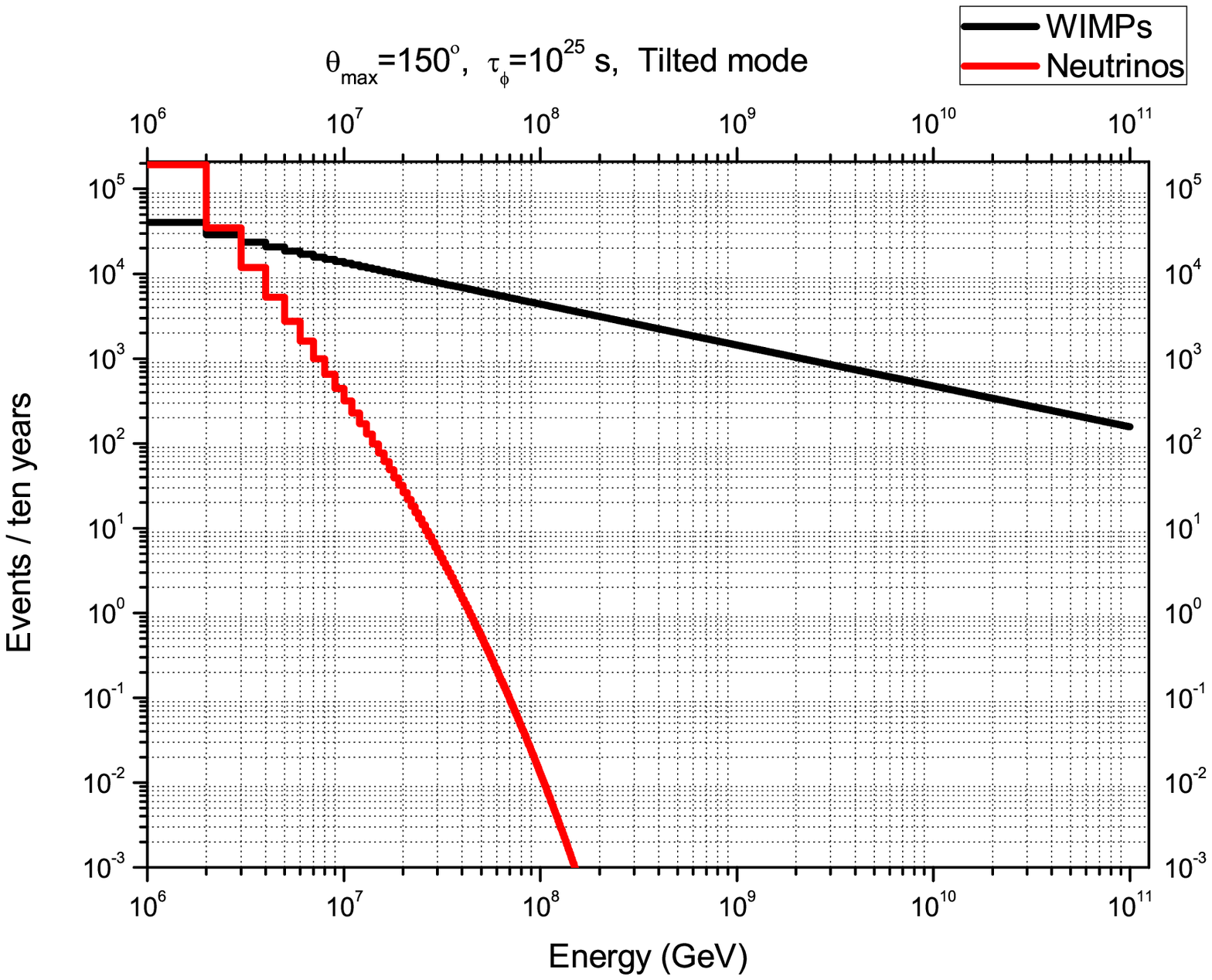}
 \caption{The WIMP and neutrino event rates are evaluated at $\theta_{max} = 150^{\circ}$ in the tilted mode at JEM-EUSO when $\tau_{\phi}=10^{25}$s}
 \label{fig:tilt_150_25}
\end{figure}

\begin{figure}
 \centering
 \includegraphics[width=0.9\textwidth]{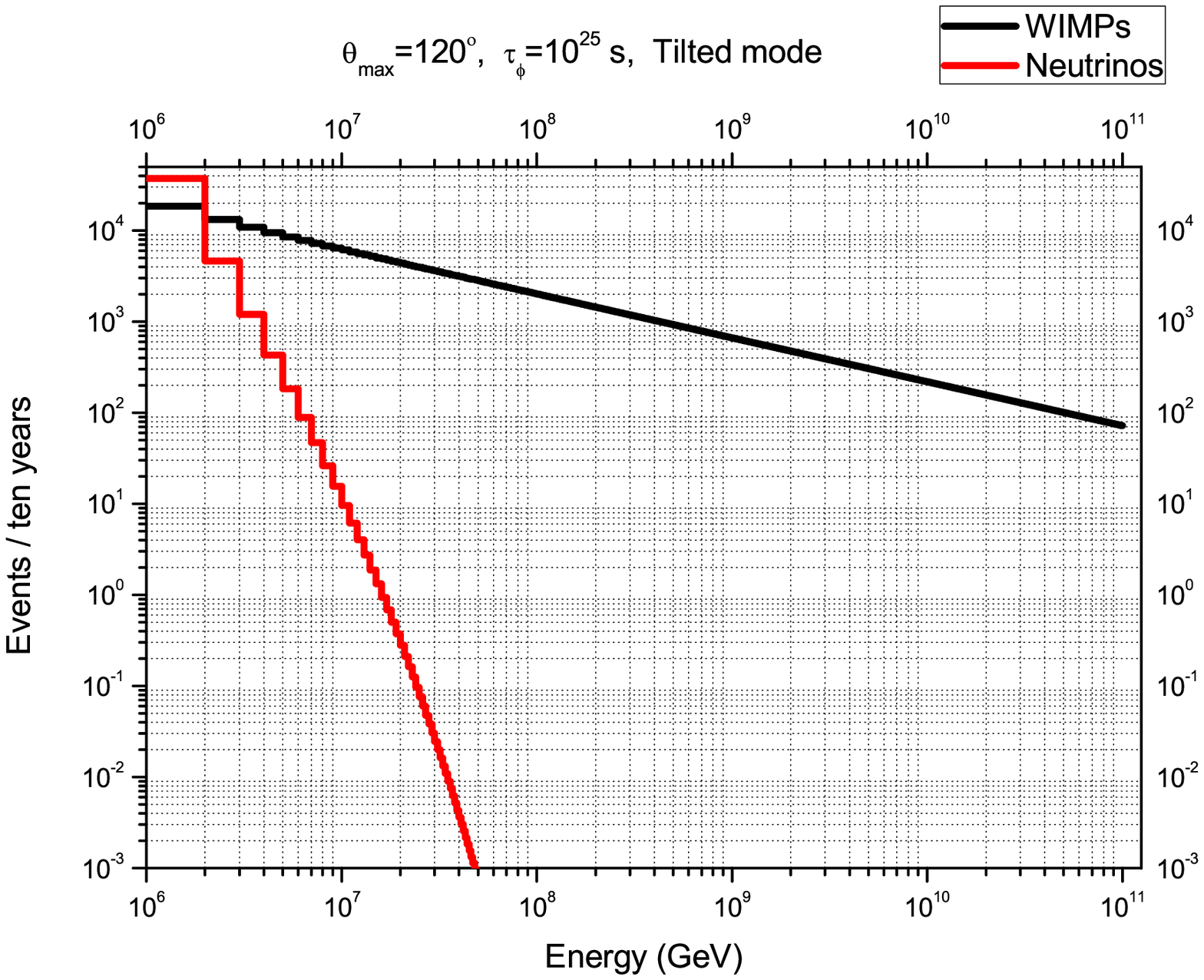}
 \caption{The WIMP and neutrino event rates are evaluated at $\theta_{max} = 120^{\circ}$ in the tilted mode at JEM-EUSO when $\tau_{\phi}=10^{25}$s}
 \label{fig:tilt_120_25}
\end{figure}

\begin{figure}
 \centering
 \includegraphics[width=0.9\textwidth]{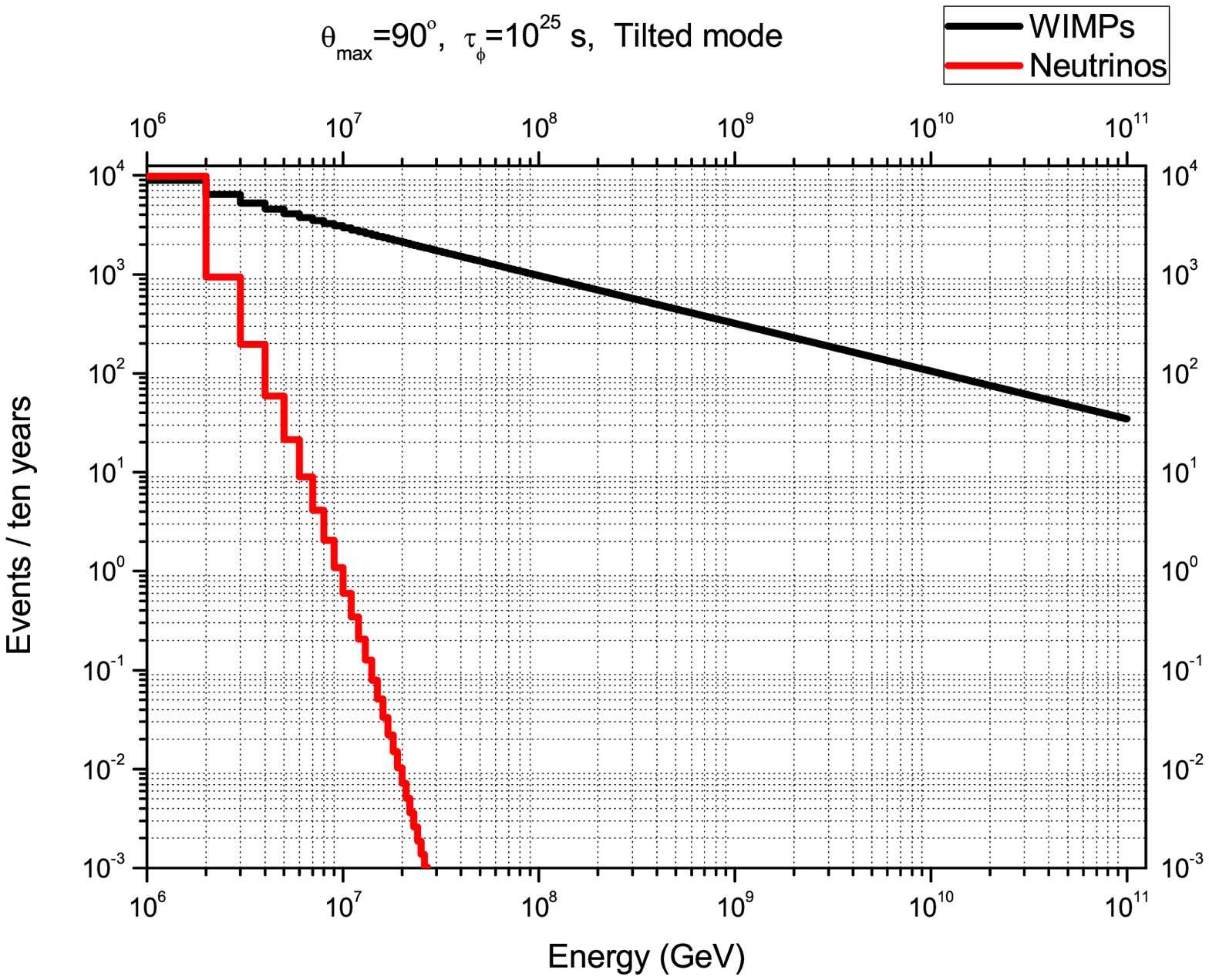}
 \caption{The WIMP and neutrino event rates are evaluated at $\theta_{max} = 90^{\circ}$ in the tilted mode at JEM-EUSO when $\tau_{\phi}=10^{25}$s}
 \label{fig:tilt_90_25}
\end{figure}

\end{document}